\documentclass{article} 
\usepackage{pstricks}

\usepackage{amsfonts,amssymb}
\usepackage{stmaryrd}
\usepackage{amsmath}

\newcommand{\AAA}{{\cal A}}

\newcommand{\CCC}{{\cal C}}

\newcommand{\EEE}{{\cal E}}
\newcommand{\FFF}{{\cal F}}
\newcommand{\GGG}{{\cal G}}

\newcommand{\OOO}{{\cal O}}

\newcommand{\QQQ}{{\cal Q}}
\newcommand{\RRR}{{\cal R}}

\newcommand{\TTT}{{\cal T}}
\newcommand{\UUU}{{\cal U}}
\newcommand{\VVV}{{\cal V}}

\newcommand{\XXX}{{\cal X}}

\renewcommand{\Bbb}{\mathbb}

\newcommand{\NNn}{{\Bbb N}}

\newcommand{\TTt}{{\Bbb T}}

\newcommand{\ZZz}{{\Bbb Z}}

\newcounter{countroman}
{\begin{list}{{\rm (\roman{countroman})}}{\usecounter{countroman}}}%
{\end{list}}

\newcounter{countalpha}
\newenvironment{anumerate}%
{\begin{list}{(\alph{countalpha})}{\usecounter{countalpha}}}%
{\end{list}}

\newcounter{countalphabf}
{\protect\begin{list}{{\rm (}{\bf \protect\alph{countalphabf}}{\rm%
)}}{\protect\usecounter{countalphabf}}}%
{\end{list}}

\newcommand{\bitmz}{\vspace{-.5\baselineskip}\begin{itemize}}
\newcommand{\eitmz}{\end{itemize}\vspace{-.25\baselineskip}}

\newcommand{\bdesc}{\vspace{-.5\baselineskip}\begin{description}}
\newcommand{\edesc}{\end{description}\vspace{-.25\baselineskip}}

\mathcode`\<="4268 
\mathcode`\>="5269 
\mathchardef\gt="313E 
\mathchardef\lt="313C 
 \def\pushright#1{{
    \parfillskip=0pt            
    \widowpenalty=10000         
    \displaywidowpenalty=10000  
    \finalhyphendemerits=0      
    \leavevmode                 
    \unskip                     
    \nobreak                    
    \hfil                       
    \penalty50                  
    \hskip.2em                  
    \null                       
    \hfill                      
    {#1}                        
    \par}}                      

 \def\qed{\pushright{$\square$}\penalty-700 \smallskip}

\newenvironment{prf}[1]{\begin{trivlist} \item[{\bf ~Proof}#1.]}%
{\qed\end{trivlist}}

\newcommand{\be}[1]{\begin{#1}}

\newcommand{\ee}[1]{\end{#1}}
\newcommand{\beq}{\begin{equation}}
\newcommand{\eeq}{\end{equation}}
\newcommand{\ba}[1]{\begin{array}{#1}}
\newcommand{\ea}{\end{array}}
\newcommand{\bea}{\begin{eqnarray}}
\newcommand{\eea}{\end{eqnarray}}
\newcommand{\bear}{\begin{eqnarray*}}
\newcommand{\eear}{\end{eqnarray*}}
\newcommand{\bpr}{\begin{prf}{}}
\newcommand{\epr}{\end{prf}}
\newcommand{\bprf}[1]{\begin{prf}{#1}}
\newcommand{\eprf}{\end{prf}}

\newtheorem{thm}{Theorem}[section]
\newtheorem{defn}[thm]{Definition}

\newtheorem{prop}[thm]{Proposition}
\newtheorem{lemma}[thm]{Lemma}

\newtheorem{cond}{}[thm]

\newtheorem{prenumb}[thm]{\hspace{-1ex}}

\renewcommand{\to}{\longrightarrow}

\newcommand{\rfun}{\stackrel{\RRR}{\to}}
\newcommand{\pfn}{\rightharpoonup}

\newcommand{\rf}{f}
\newcommand{\rg}{g}

\newcommand{\cod}[1]{
\llbracket{#1}\rrbracket
}

\newcommand{\rar}{\triangleright}

\newcommand{\orig}[1]{\overrightarrow{#1}}
\newcommand{\tagg}[1]{\tag{\sf{#1}}}
\newcommand{\Prob}{{\rm Prob}}
\newcommand{\Var}{\XXX}

\newcommand{\guardss}[3]{{#1}\ \ \mbox{ {\sf \small guards}}\ \  {#2}\ \ \mbox{ {\sf \small within}}\ \ {#3}}

\newcommand{\pder}[2]{\big[#1\ {\mbox{\large ${\vdash}$}}\ #2\big]}
\newcommand{\ppder}[1]{\big[#1\big]}

\newcommand{\adv}[2]{{\sf Adv}\big[#1\ {\mbox{\large ${\vdash}$}}\ #2\big]}

\newcommand{\indep}[2]{\big[#1\thinspace \bot\thinspace #2\big]}

\newcommand{\crbefo}[2]{ #1({#2})}

\newcommand{\subs}[2]{{#1}^{\wcard{#2}}}

\newcommand{\wcard}{\circledast}
\newcommand{\dom}{{\sf dom}}
\newcommand{\kernel}{\kappa}
\newcommand{\token}{h}
\newcommand{\startt}{\underline\tau}
\newcommand{\stopp}{\overline\tau}

\newcommand{\sstr}[1]{{#1}^\omega}
\newcommand{\Randf}{\RRR}
\newcommand{\Feasf}{\FFF}
\renewcommand{\paragraph}[1]{\smallskip\noindent{\bf #1}}
\newcommand{\Rand}{\widetilde\RRR}

\newcommand{\FV}[1]{\Var({#1})}
\newcommand{\Qneg}{\smash{\frac{1}{Q}}}

\begin{document}
  
  \title{Quantifying pervasive authentication:\\
the case of the Hancke-Kuhn protocol} 
  \author{Dusko Pavlovic\thanks{Kestrel Institute and Oxford University} and Catherine Meadows\thanks{Naval Research Laboratory}}
\date{}

\maketitle

\begin{abstract}
As mobile devices pervade physical space, the familiar authentication patterns are becoming insufficient: besides entity authentication, many applications require, e.g., location authentication. Many interesting protocols have been proposed and implemented to provide such strengthened forms of authentication, but there are very few proofs that such protocols satisfy the required security properties. In some cases, the proofs can be provided in the symbolic model. More often, various physical factors invalidate the perfect cryptography assumption, and the symbolic model does not apply. In such cases, the protocol cannot be secure in an absolute logical sense, but only with a high probability.  But while probabilistic reasoning is thus necessary, the analysis in the full computational model may not be warranted, since the protocol security does not depend on any computational assumptions, or on attacker's computational power, but only on some guessing chances.

We refine the Dolev-Yao algebraic method for protocol analysis by a probabilistic model of guessing, needed to analyze protocols that mix weak cryptography with physical properties of nonstandard communication channels. Applying this model, we provide a precise security proof for a proximity authentication protocol, due to Hancke and Kuhn, that uses probabilistic reasoning to achieve its goals.
\end{abstract}

\section{Introduction}
\paragraph{Two paradigms of security.}
Traditionally, two paradigms have been used for proving protocol security.  The first one, captured by the symbolic model, commonly known as ``Dolev-Yao", describes both protocol and attacker in terms of an algebraic theory \cite{Dolev-Yao}.  While this has been criticized as crude, it is often highly effective and easily automated.  The other paradigm, captured by the computational model, usually relies on some notion of indistinguishability from the point of view of a computationally limited attacker \cite{GoldreichI}. Recently, a lot of research \cite{BackesM:COSP,CortierV:SymbSurvey}, starting with \cite{Abadi-Rogaway}, has been devoted to drawing the two paradigms closer together.  This strategy has generally been to rely upon crypto-algorithms that themselves satisfy strong enough definitions of security, so that, if used in the proper way, they can be treated as Dolev-Yao ``black boxes''.

\paragraph{Problem of pervasive security.} However, there is an emerging class of security protocols for which it seems difficult to bring these two paradigms together. Such protocols arise in heterogenous networks of diverse computational and communication devices, with mixed type channels between them \cite{Wong-Stajano07}. Nowadays ubiquitous, such networks can be viewed as a realization of Doug Engelbart's visionary idea of {\em smart space}\/ and {\em pervasive computation} \cite{EngelbartD:Augmenting}. The spatial aspects of computation give rise to a new family of security problems, where the standard authentication requirements need to be strengthened by proofs of spatial proximity. In some cases, it has been possible to refine symbolic methods to get stronger proofs \cite{PavlovicD:dist06,Schaller09}. But there are other cases that resist symbolic analysis.
One such case is the {\em Hancke-Kuhn distance bounding protocol\/}  \cite{Hancke-Kuhn}, which we analyze in the present paper.
The protocol consists of a timed challenge-response exchange in which a prover Peggy needs to convince a verifier Victor that she is in the vicinity.  Peggy's rapid response to Victor's challenge is implemented using a rapidly computable function. The requirement that the function must be rapidly computable turns out to weaken it cryptographically. One of the main requirements of cryptographic strength is {\em diffusion}: for a boolean function, each bit of the output should depend on each bit of the input \cite{Webster-Tavares}. But a function that has to wait for the last bit of its input before it produces the first bit of its output is not rapidly computable. The other way around, an {\em on-line\/} function, that produces its output while still receiving its input, is easier to compute, but cannot be cryptographically strong. So there is a tradeoff between cryptographic strength and rapid computability. We explore this tradeoff in Sec.~\ref{boxplus}, and quantify the information leakage of on-line functions. The Hancke-Kuhn protocol is based on such a function.

Already in the original presentation \cite{Hancke-Kuhn} of their protocol, Hancke and Kuhn wrote down an estimate of the attacker's chance to guess a response bit. However, besides attempting to guess some bits of the response, the attacker may also attempt to guess the secret on which the response is based. Moreover, he may attempt his guesses directly, or make use of the responses stored from other sessions. Last but not least, he may collude with Peggy. Towards a precise security proof, the diverse strategies available to the attacker must be evaluated together, and exhaustively. This requires a formal model of protocol execution. 

\paragraph{Bayesian security.} But what model to use? The symbolic model cannot be used because the perfect cryptography assumption is not validated by the on-line function, which is the central feature of the protocol. On the other hand, the cryptographic strength and weakness of this function, and the resulting security and insecurity of their protocol, does not have anything to do with any computational assumptions, or with the computational power of the adversary: it only depends on guessing chances, which cannot be essentially increased by computational power. Thus using the computational model does not contribute to the analysis of the central feature of the protocol, although it does apply to any implementation.

The most natural model for analyzing the Hancke-Kuhn protocol that we came up with extends the symbolic model by a rudimentary probabilistic theory of guessing. 
It retains the perfect cryptography assumption for the standard cryptographic primitives used in the protocol, in particular for the keyed hash function. In a probabilistic context, though, the perfect cryptography assumption means that the output distributions of the relevant cryptographic primitives are statistically indistinguishable from the uniform distribution. Assuming this for the hash function used in the protocol brings us close to the {\em random oracle assumption}, often used in computational analyses \cite{BellareM:oracles}. There is a sense in which  the random oracle assumption can be construed as the probabilistic version of the perfect cryptography assumption.

In summary, we contend that the simplest model capturing the central features of the Hancke-Kuhn authentication protocol must be probabilistic, but need not be computational. The probabilistic model that we propose is an extension of the symbolic theories used in our previous work \cite{PavlovicD:ESORICS04,PavlovicD:CSFW05,PavlovicD:ESORICS06}. On the other hand, a version of the standard computational model can be obtained as an extension of this probabilistic model (by distinguishing a submonoid of feasible functions within our monoid of randomized boolean functions). It should be noted that these logical maps between the models go in the opposite direction from those in the explorations of the computational soundness of the various fragments of the symbolic model \cite{Abadi-Rogaway,BackesM:COSP,CortierV:SymbSurvey}. In such explorations, the symbolic languages are mapped (interpreted) {\em in\/} the computational language; here, a more concrete model is mapped onto a more abstract model, which is its quotient, just like blocks of low-level code are mapped onto the expressions of a high-level programming language, or like more concrete state machines are mapped on more abstract state machines \cite{PavlovicD:ASE01,PavlovicD:AMAST02}. It follows that anything proven about the abstract model remains valid about its more concrete implementations:
e.g., the Bayesian reasoning about secrecy remains valid in the computational model --- provided that the assumed randomness of the hash function can be validated. 
This proviso is, of course, not satisfied in practice, since cryptographic hash functions are not truly random. The task, thus, remains to strengthen or refine the reasoning as to be able to discharge such unrealistic assumptions. This logical strategy was discusssed in \cite{PavlovicD:ESORICS04,PavlovicD:CSFW05}. While not widely accepted in security, this is a standard approach to refinement based software development: e.g., Euclid's algorithm is usually described assuming the ring of integers; but the assumption that there are infinitely many integers must be discharged before the algorithm is implemented in a real computer.

The space does not allow us to delve into the details of this approach, as applied to security. They will be presented elsewhere. In the present paper, we attempt to present a very special instance of this approach, where a modest probabilistic extension of the symbolic model suffices for the problem at hand --- yet it leads to an essentially different reasoning framework, with bayesian derivations instead of logical. The resulting technical divergence, mitigated by the conceptual guidance from the underlying simpler model, should be viewed as one of the main features of the incremental approach, pursued in the Protocol Derivation Logic (PDL)
\cite{PavlovicD:ESORICS04,PavlovicD:CSFW05,PavlovicD:ESORICS06}. In \cite{PavlovicD:dist06}, PDL was already used to analyze distance bounding protocols, similar to Hancke-Kuhn's, and for reasoning about pervasive security in general. An interesting feature of the current probabilistic extension of PDL is that the concept of {\em guards}, originally developed for reasoning about secrecy \cite{PavlovicD:ESORICS06}, now provides a crucial stepping stone into our analysis of guessing chances, and of the concrete authentication guarantees in the Hancke-Kuhn protocol in Sec.~\ref{proof}, as well as in the abstract view of symbolic authentication in Thm.~\ref{CRprop}.

%

%


\paragraph{Related work.} As already mentioned, the closest relative of the PDL formalism, underlying this work, and briefly summarized in Sec.~\ref{PDL}, is PCL \cite{PavlovicD:JCS04,PavlovicD:JCS05,PCL07}. Both formalisms owe a lot to strand spaces \cite{strands}, in spirit, and in execution models, although the logical methods diverge. Our probabilistic extension of PDL is predated by the probabilistic extension of PCL in \cite{PCL-prob}, and by the probabilistic extension of strand spaces in \cite{strands-prob}. But each of the three probabilistic approaches has a different intent, and a completely different implementation, conceptually and technically. It would be interesting to explore these differences more closely, as some tasks may yield to combined modeling methods.

\paragraph{Paper outline.}  The paper continues with a review of distance bounding authentication, and a description of the Hancke-Kuhn protocol. In Sec.~\ref{PDL} we provide a brief overview of the derivational method of protocol analysis, and of PDL. We also recall the algebraic notions of derivability and guards, originally used for derivational analyses of secrecy, and here adapted for authenticity. The probabilistic versions of these notions are introduced in Sec.~\ref{probguards}, and then used to model guessing. The gathered tools  are then put to use. In Sec.~\ref{boxplus}, we analyze the information leakage of on-line functions in general, and characterize the Hancke-Kuhn function among them. In Sec.~\ref{proof}, we quantify the authentication achieved in the Hancke-Kuhn protocol. Sec.~\ref{conclusion} closes the paper with a summary of the results and a discussion of the extensions. All proofs are in the Appendix.

\section{The Hancke-Kuhn protocol}\label{hancke}
\subsection{Background}
In a man-in-the-middle attack on a challenge-response protocol, the attacker relays messages, sometimes modified, between the legitimate participants. If resending a message takes time, the legitimate participants may observe slower traffic. This has been proposed as a method to prevent  man-in-the-middle attacks. In particular, the challenger can measure the presumed round trip of his challenge and of responder's response, and compute a maximal distance of the responder, assuming an upper bound on the message velocity. This can assure the authenticity of the response, if it is known that the attacker cannot be too close. 
This is the idea of {\em distance bounding} \cite{desmedt,Beth-Desmedt}. The early security analyses of distance bounding protocols go back to the early 1990s \cite{Brands-Chaum}. The interest in this type of authentication re-emerged recently, with the task of device pairing and a genuine need for proximity authentication in pervasive networks \cite[etc.]{sastry,capkun,clulow,Hancke-Kuhn,rope,PavlovicD:dist06,singlee,chandran09}. 
From the outset, the basic idea of distance bounding was to combine some cryptographic authentication tools, such as hashes or signatures, with a physical constraint, such as the limited speed of message exchange. Most distance bounding protocols \cite{Brands-Chaum,capkun,PavlovicD:dist06} implement this combination by using two channel types: the standard network channels for the cryptographic authentication, and  the timed channels for the rapid response. The Hancke-Kuhn protocol \cite{Hancke-Kuhn} stands out by it simplicity, and by the fact that both cryptographic data and the rapid response are sent on the timed channel. This, however, comes for the price of information leakage, which makes the security analysis interesting.

\subsection{The protocol}
As mentioned before, the goal of the Hancke-Kuhn protocol is that the prover Peggy proves to the verifier Victor that she is nearby. It is assumed that Peggy and Victor share a long term secret $s$, and a public hash function $H$. The relevant security requirement from $H$ will turn out to be a version of the range preimage resistance \cite{Rogaway-Shrimpton}. The simplest way to present a protocol session is to view it in two stages. 

In the {\em first stage}, Peggy and Victor exchange values $a$ and $b$, which can be predictable for the attacker, but must never be reused by Peggy and Victor in more than one protocol session. The values $a$ and $b$ can thus be viewed as counters.

\newcommand{\alice}{V}
\newcommand{\bob}{P}
\newcommand{\newa}{\scriptstyle \nu x}
\newcommand{\msgOne}{\scriptstyle \startt
 <x>}
\newcommand{\msgOneBack}{\scriptstyle \stopp\left(x \boxplus {\token}\right)}
\newcommand{\intoOne}{\scriptstyle (x)}
\newcommand{\intoOneBack}{\scriptstyle \left<x \boxplus {\token}\right>}
\begin{figure}[htbp]
\begin{center}
\def\JPicScale{0.6}
\ifx\JPicScale\undefined\def\JPicScale{1}\fi
\psset{unit=\JPicScale mm}
\psset{linewidth=0.3,dotsep=1,hatchwidth=0.3,hatchsep=1.5,shadowsize=1,dimen=middle}
\psset{dotsize=0.7 2.5,dotscale=1 1,fillcolor=black}
\psset{arrowsize=1 2,arrowlength=1,arrowinset=0.25,tbarsize=0.7 5,bracketlength=0.15,rbracketlength=0.15}
\begin{pspicture}(0,0)(102.5,70)
\newrgbcolor{userLineColour}{0.4 0.4 0.4}
\pspolygon[linewidth=0.2,linecolor=userLineColour](-2.5,70)(32.5,70)(32.5,0)(-2.5,0)
\newrgbcolor{userLineColour}{0.4 0.4 0.4}
\pspolygon[linewidth=0.2,linecolor=userLineColour,linestyle=dashed,dash=1 1](67.5,70)(102.5,70)(102.5,0)(67.5,0)
\psline[border=0.75,arrowscale=1 1.45,arrowsize=1.45 2,arrowlength=1.55]{->}(21.88,47.5)(80.62,42.5)
\psline[border=0.75,arrowscale=1 1.45,arrowsize=1.45 2,arrowlength=1.55]{->}(76.88,17.5)(23.75,12.5)
\psline[arrowsize=1.45 2,arrowlength=1.55]{->}(15,57.5)(15,50.62)
\psline[arrowsize=1.45 2,arrowlength=1.55]{->}(85,40)(85,21.25)
\rput[l](2.5,65){$\alice$}
\rput[l](72.5,65){$\bob$}
\rput(15,60){$\newa$}
\rput(15,47.5){$\msgOne$}
\rput(15,12.5){$\msgOneBack$}
\rput[b](22.5,52.5){}
\rput(85,42.5){$\intoOne$}
\rput(85,17.5){$\intoOneBack$}
\psline[arrowsize=1.45 2,arrowlength=1.55]{->}(15,45)(15,16.25)
\end{pspicture}
\caption{Hancke-Kuhn protocol: Second Stage}
\label{HK}
\end{center}
\end{figure}
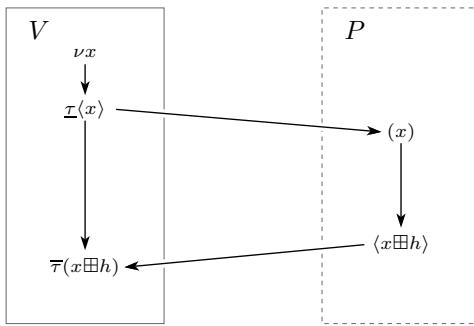
In the {\em second stage}, Peggy and Victor both form the hash ${\token} = H(s::a::b)$ and proceed with the exchange on Fig.~\ref{HK}. If Victor's challenge $x = (x_i)\in \ZZz_2^\ell$ is a bitstring of length $\ell$, then the hash $\token$ should be $2\ell$ bits long which we view as a concatenation $\token = \token^{(0)}::\token^{(1)}\in \ZZz_2^{2\ell}$ of two strings of $\ell$ bits. The function $\boxplus : \ZZz_2^\ell \times \ZZz_2^{2\ell} \to \ZZz_2^\ell$ is defined bitwise for $i=1,2,\ldots, \ell$ by
\bea\label{HKfun}
(x \boxplus h)_i & = & h_i^{(x_i)}
\eea
To summarize Fig.~\ref{HK},
\begin{itemize}
\item Victor generates a random bitstring $x$ of length $\ell$, and sends each bit $x_i$ of $x$ at times $\startt_i$. 

\item To each bit $x_i$, Peggy responds with $h^{(0)}_i$ if $x_i=0$, and with $h^{(1)}_i$ if $x_i = 1$.

\item Victor receives Peggy's $i$-th bit response at time $\stopp_i$. He knows $\token$ as well, and can check that these responses are correct. If only he and Peggy know $h$, then the responder must be Peggy. 
He then uses the times between the sending the challenges and receiving the responses, together with the velocity of the message signal, to compute his distance from Peggy.
\end{itemize}

\subsection{Discussion}
\paragraph{Leaking information to the attacker.} The crucial component of the protocol is the Hancke-Kuhn function $\boxplus$. Its main feature is that it is rapidly computable, as efficiently as the {\em exclusive or} $\oplus$. It is thus as suitable for timed authentication as $\oplus$, but it also leaks information, although less than $\oplus$: while $x$ and $x\oplus g$ allow extracting $g$ because $g = x\oplus x\oplus g$, $x$ and $x\boxplus \token$ allow extracting only half of the bits of $\token$. However, it is easy to see from \eqref{HKfun} that from $x$, and $x\boxplus \token$, {\em and moreover\/} $(\neg x) \boxplus \token$, the attacker can extract all of $\token$. That is why Peggy and Victor must not reuse their counters. If $\token = H(s::a::b)$ can be used in two responses, then an attacker can challenge Peggy twice, first with $x$ and then with $\neg x$, and thus get $x\boxplus \token$ and $(\neg x) \boxplus \token$ as the two responses. From this, he can extract $\token$ and impersonate Peggy to Victor. Even if the counters are never reused, the fact that half of the response bits can be acquired by an attacker needs to be carefully examined, and his chances to guess the rest evaluated.

\paragraph{Overlooked assumption.} Hancke and Kuhn's estimate that the probability that an attacker may succeed in impersonating Peggy is $(\frac{3}{4})^{|x|}$  relies on the implicit assumption that $|x| \leq |s|$.  Otherwise, if $|x| \gt |s|$, the attacker has better odds to guess $s$ than $x$. In practice, of course, the assumption $|x| \leq |s|$ is usually satisfied, because the secret $s$ is usually at least 256 bits long, while the challenge $x$ may be shorter. Strictly speaking, though, the impression that protocol's security only depends on the length of the challenge $x$ is not correct, since a short secret $s$ would make it vulnerable. 

\paragraph{Dishonest prover and the kernel.} Another interesting weakness is that the value of Peggy's $i$-th response bit $(x \boxplus {\token})_i$ does not depend on $x_i$ if ${\token}^{(0)}_i = {\token}^{(1)}_{i}$. A dishonest Peggy can thus analyze the hash $\token$ and respond without waiting for $x_i$ whenever ${\token}^{(0)}_i = {\token}^{(1)}_{i}$. If the response time is averaged, she is likely to appear closer to Victor than she really is. 

Since Victor's counter $b$ is predictable, Peggy can attempt to choose her own counter $a$ to maximize the size of the {\em kernel\/} $\kernel \token$ of $\token= H(s::a::b)$, defined
\bea\label{kernell}
\kernel \token & = & \{i\leq \ell\ |\ \token^{(0)}_i = \token^{(1)}_i \}
\eea
The larger the kernel, the closer Peggy can appear to Victor. However, the problem of finding a value $a$ such that, for a fixed $s$ and $b$, the image $H(s::a::b)$ has a desired property is a  version of the range preimage problem \cite{Rogaway-Shrimpton}. The assumption that $H$ is a hash function, and in particular that it is a one-way function, implies that dishonest Peggy's advantage in finding a preimage $a$ such that $H(s::a::b)$, given $s$ and $b$, falls within a desired range of strings with a large kernel, is negligible. This means that dishonest prover's manipulation of the kernel is unfeasible.

Further ad hoc observations get more complicated, without providing any definite assurances. This demonstrates the need for a rigorous analysis within a formal model.

\paragraph{Modeling the essence of the Hancke-Kuhn protocol.} The assumption that $H$ is a one-way function will turn out to be the only point where the security of the Hancke-Kuhn protocol depends on computation. All other attack strategies only involve guessing chances. To show this, in the following sections we introduce a probabilistic (Bayesian) protocol model, which strictly extends the standard algebraic (symbolic) model, and is a strict fragment of the standard computational model. The hash $H$ is modeled as a randomized function, as defined in Sec.~\ref{probguards}. The perfect cryptography assumption of the symbolic model lifts in our Bayesian model to the  assumption that the hashes are truly random, which is, of course, analogous to the {\em random oracle\/} assumption in the computational model. It allows us to abstract away the generic and negligible vulnerabilities, and to focus on the interesting aspects of the security of the Hancke-Kuhn protocol, achieved {\em in spite\/} of the cryptographic weakness of the $\boxplus$ function as it central feature.

\section{Algebraic protocol models}\label{PDL}
We analyze the Hancke-Kuhn protocol by the derivational method. The varied versions of this method have been applied to many protocols
\cite{PavlovicD:JCS04,PavlovicD:ESORICS04,PavlovicD:CSFW05,PavlovicD:JCS05,PCL07}. While the algebraic protocol model suffices in most cases, the Hancke-Kuhn protocol requires an evaluation of guessing chances. We attempt to find a simple model that will allow this.

\subsection{Message algebras}\label{MesAlg}
In the Dolev-Yao protocol model, messages are represented as terms of a free algebra of encryption and decryption operations  \cite{Dolev-Yao}. More general algebraic models allow additional operations, and additional equations \cite{CortierV:algebraic}. Recall that an algebraic theory is a pair $(O,E)$, where $O$ is a set of finitary operations (given as symbols with arities), and $E$ a set of well-formed equations (i.e. where each operation has a correct number of arguments) \cite{Gratzer}.

\be{defn}\label{mesalgdef}
An algebraic theory $\TTt = (O,E)$ is called a {\em message theory\/} if $O$ includes 
a binary operation of pairing $<-,->$, and
the unary operations $\pi_1$ and $\pi_2$, 
such that $E$ contains the equations 
$\pi_1(u,v) =  u$, $\pi_2(u,v) =v$, and
$\left(\left(x,y\right),z\right)  = \left(x,\left(y,z\right)\right)$.
A {\em message algebra} is a polynomial extension $\TTT[\Var]$ of a $\TTt$-algebra $\TTT$.
\ee{defn} 

\paragraph{Remarks.} The third equation implies that there is a unique $n$-tupling operation for every $n$. The first two imply that the components of any tuple can be recovered. 
A polynomial extension $\TTT[\Var]$ is the free $\TTt$-algebra generated by adjoining a set of {\em indeterminates\/} $\Var$ to a $\TTt$-algebra $\TTT$ \cite[\S8]{Gratzer}. The elements $x,y,z\ldots$ of $\Var$ are used to represent nonces and other randomly generated values. This is justified by the fact that indeterminates can be consistently renamed: nothing changes if we permute them. That is just the property required from the random values  generated in a run of a protocol\footnote{Of course, this is not the only requirement imposed on nonces and random values. The other requirement is that they are known only locally, i.e. by those principals who generate them, or who receive them unencrypted. This requirement is not formalized within the algebra of messages, but by the binding rules of process calculus or actions by which the messages are sent \cite{PavlovicD:JCS05,PavlovicD:ESORICS06}.}.

\subsection{Protocol models}\label{ProtocolModels}
There are several protocol modeling formalisms that can be used for protocol derivations. The process calculus in \cite{PavlovicD:JCS04,PavlovicD:JCS05} was designed specifically for this purpose. Strand spaces \cite{strands} were designed for a different purpose, but they can be adapted for protocol derivations too. In \cite{PavlovicD:ESORICS04,PavlovicD:CSFW05,PavlovicD:ESORICS06} we used partially ordered multisets (pomsets) of actions \cite{Pratt:Pomsets}, which allow simple tool support \cite{PavlovicD:ARSPA06}. We stick with this approach, but the subtle (or in some cases not so subtle) differences between these approaches are of no consequence here. For completeness, we provide a brief overview. For more detail, the reader may want to consult some of the mentioned references.

In all cases, the set of actions $\AAA$ is generated over the message algebra $\TTT[\Var]$ by a grammar allowing each term $t\in \TTT[\Var]$ to be sent in the action $<t>\in \AAA$, and received in the action $(t)\in \AAA$. Moreover, an indeterminate $x\in \Var$ can be introduced into a protocol by the binding action $(\nu x)\in\AAA$, which is read as "generate fresh $x$". 

\subsubsection*{Challenge-response}
\renewcommand{\msgOne}{\scriptstyle <c^{VP}x>}
\renewcommand{\msgOneBack}{\scriptstyle (r^{VP}x)}
\renewcommand{\intoOne}{\scriptstyle ((c^{VP}x))}
\renewcommand{\intoOneBack}{\scriptstyle <<r^{VP}x>>}
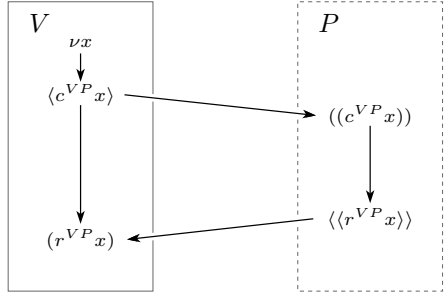
\begin{figure}[htbp]
\begin{center}
\def\JPicScale{0.55}
\ifx\JPicScale\undefined\def\JPicScale{1}\fi
\psset{unit=\JPicScale mm}
\psset{linewidth=0.3,dotsep=1,hatchwidth=0.3,hatchsep=1.5,shadowsize=1,dimen=middle}
\psset{dotsize=0.7 2.5,dotscale=1 1,fillcolor=black}
\psset{arrowsize=1 2,arrowlength=1,arrowinset=0.25,tbarsize=0.7 5,bracketlength=0.15,rbracketlength=0.15}
\begin{pspicture}(0,0)(102.5,70)
\newrgbcolor{userLineColour}{0.4 0.4 0.4}
\pspolygon[linewidth=0.2,linecolor=userLineColour](-2.5,70)(32.5,70)(32.5,0)(-2.5,0)
\newrgbcolor{userLineColour}{0.4 0.4 0.4}
\pspolygon[linewidth=0.2,linecolor=userLineColour,linestyle=dashed,dash=1 1](67.5,70)(102.5,70)(102.5,0)(67.5,0)
\psline[border=0.75,arrowscale=1 1.45,arrowsize=1.45 2,arrowlength=1.55]{->}(25.62,47.5)(71.88,42.5)
\psline[border=0.75,arrowscale=1 1.45,arrowsize=1.45 2,arrowlength=1.55]{->}(71.25,17.5)(26.25,12.5)
\psline[arrowsize=1.45 2,arrowlength=1.55]{->}(15,57.5)(15,50.62)
\psline[arrowsize=1.45 2,arrowlength=1.55]{->}(85,40)(85,21.25)
\rput[l](2.5,65){$\alice$}
\rput[l](72.5,65){$\bob$}
\rput(15,60){$\newa$}
\rput(15,47.5){$\msgOne$}
\rput(15,12.5){$\msgOneBack$}
\rput[b](22.5,52.5){}
\rput(85,42.5){$\intoOne$}
\rput(85,17.5){$\intoOneBack$}
\psline[arrowsize=1.45 2,arrowlength=1.55]{->}(15,45)(15,16.25)
\end{pspicture}
\caption{CR template}
\label{CR}
\end{center}
\end{figure}
Fig.~\ref{CR} shows the abstract challenge-response protocol template, where the verifier $V$ictor authenticates the prover $P$eggy. It is assumed that only Peggy is able to transform the fresh challenge $c^{VP} x$ into the response $r^{VP} x$. This assumption is construed as a constraint on the operations $c^{VP}$ and $r^{VP}$. The actions $<<t>>$, and $((t))$ are syntactic sugar for ``send (resp. receive) a message from which anyone can extract $t$".

\subsection{Views, derivability and guards}\label{States}
As usual, the communication channels are assumed to be controlled by the attacker: she observes all sent messages, and controls their delivery. However, she may not be able to invert all operations, and she has no insight into the fresh or secret data of other principals. Hence the different {\em views\/} of the various protocol participants.

A state $\sigma$ reached in a protocol execution is a lower closed pomset of actions executed up to that point, with an assignment of values to principals' local variables, which they use to store messages and their local computations. The view $\Gamma_P^\sigma$ of a principal $P$ at a state $\sigma$ consists of all terms that $P$ may have observed up to $\sigma$, and all terms that she could derive from that. Formally, this last clause means that $\Gamma_P^\sigma$ is upper closed under the derivability relation
\be{multline}\label{algder}
\Xi\ \vdash  \Theta \ \  \iff    \forall t \in \Theta\ \exists \varphi\in O^{(n)}\ \exists s_1,\ldots,s_n \in \Xi.\  t \stackrel{E} = \varphi(s_1,\ldots,s_n)
\end{multline}

\noindent where $\Xi,\Theta\subseteq \TTT[\Var]$ are finite sets of terms, $O^{(n)}$ is the set of well-formed $n$-ary operations in the signature $O$, and the equation is derivable from $E$. 

\subsubsection*{Authentication by challenge-response}
The challenge-response protocol in Fig.~\ref{CR} validates authentication if Victor is justified in drawing a global conclusion from his local observation: i.e., having observed his own actions in  on the left, Victor should have good reasons to conclude that Peggy must have performed her actions on the right, and that all these actions should be ordered as on the figure. Intuitively, this conclusion of Victor's can be justified by the assumptions that
 \begin{enumerate}
  \item anyone who originated the response $r^{VP}x$ had to previously receive the challenge $c^{VP}x$, which could only happen after Victor sent this challenge;
 \item no one could produce $r^{VP}x$ without knowing the secret $s^{VP}$, so it must be Peggy.
 \end{enumerate}
This last conclusion is based on the assumption that only Peggy knows $s^{VP}$, or only Peggy and Victor. In both cases, Victor's reasoning is the same, because he knows that he did not send $r^{VP} x$.
 
Using the derivability relation, these informal justifications can be refined into slightly more formal proof obligations in terms of \eqref{algder}, as follows. For any set of principals $\Pi$, it is required that
\begin{enumerate}
\item whenever there is a derivation  $\Xi  \vdash  r^{VP}x$, then there must also be a derivation $\Xi \vdash c^{VP}x$, for any set of terms $\Xi$ observed by $\Pi$ in a run of CR {\em before} $r^{VP}x$ is sent;
\item whenever there is a derivation  $\Xi,c^{VP}x  \vdash  r^{VP}x$, then there must also be a derivation $\Xi,c^{VP}x \vdash s^{VP}$, for any set of terms $\Xi$ known to $\Pi$ in a run of CR  before $r^{VP}x$ is sent.
\end{enumerate}

This type of authentication reasoning can be formalized using the notion of {\em guards} from \cite{PavlovicD:ESORICS06}.

\be{defn}\label{alguards}
We say that a set of sets of terms $\GGG$ \/ {\em algebraically guards} a term $t$ with respect to a set of terms $\Upsilon$, and write ${\GGG}\ \mbox{{\sf \small guards}}\  {t}\ \mbox{{\sf \small within}}\ {\Upsilon}$ if for all $\Xi\subseteq \Upsilon holds$
\bea\label{alguardseq}\Xi \vdash t  & \Rightarrow &  \exists \Gamma\in \GGG.\ \ \Xi\vdash \Gamma 
\eea
\ee{defn}

\paragraph{Explanation.} We say that, in a context $\CCC$, $\GGG$ guards $t$ if every computation path to $t$ leads through some element of $\GGG$. In other words, if $\Xi$ allows computing $t$, then it is "because" it allows computing some of $t$'s guards from $\GGG$. 

\paragraph{Example.} Let $\Upsilon =  (DH)$ be the set of terms that may become known to the participants and eavesdroppers of a run of the Diffie-Hellman protocol. Then $$\guardss {\big\{\{x,g^y\}, \{y,g^x\} \big\}}{g^{xy}}{(DH)}$$
Note that $g^{xy}$ can be derived not only from $\{x,g^y\}$ and $\{y,g^x\}$ but also from $\{g,x,y\}$ and $\{g,xy\}$; however, neither of these sets can occur in a run of the Diffie-Hellman protocol between two honest principals, so they are not contained in the set $\Upsilon = (DH)$.

\begin{defn}\label{termcont}
Let $\QQQ$ be a protocol run, and $A$ a set of actions in $\QQQ$. The {\em term context} is the set
\bear
\crbefo \QQQ A & = & \bigcup_{P \in \Pi} \Gamma^\iota_P \cup \Gamma^{\rar A}_P
\eear
where $\Pi$ is the set of principals engaged in the run, $\Gamma^\iota_P$ is the set of terms known to a principal $P$ initially, and $\Gamma^{\rar A}_P$ is the set of terms known to $P$ {\em before} any of the actions $a\in A$ are executed in $\QQQ$.
\ee{defn}

Using the guard relation, we can prove that the challenge-response protocol validates authentication.

\be{thm}\label{CRprop} Let $\QQQ$ be a run of the challenge-response protocol on Fig.~\ref{CR}.  Suppose that the functions $c^{VP}$ and $r^{VP}$ satisfy
{\small
\[
\guardss{\left\{\{c^{VP}x, s^{VP}\}\right\}}{r^{VP} x}  { \crbefo {\QQQ }{r^{VP} x}}
\]
}
\hspace{-.4em}where $s^{VP}$ is a secret known only to Peggy (and possibly  to Victor). Then Victor is justified in drawing the following global conclusion from his local observations:
\begin{align}
V : &\ \ (\nu x)_V \rar <c^{VP}x>_V\hspace{5em} \rar  
(r^{VP}x)_{V} 
 \notag \\
\Longrightarrow & \Big((\nu x)_V \rar <c^{VP}x>_V \rar ((c^{VP}x))_P \rar 
 <<r^{VP}x>>_{\orig P}\rar
(r^{VP}x)_{V}\Big) \tagg{cr} 
\end{align}
\hspace{-.4em}where the relation $a \rar b$ says that action $a$ occurs before action $b$, and $<<m>>_{\orig P}$ denotes the first time $P$ sends message $m$ after creating it. 
\ee{thm}

The proof of this theorem is obtained by expanding the definition of the guard relation and analyzing the term context of the challenge-response protocol. Several examples of reasoning with this relation can be found in \cite{PavlovicD:ESORICS06}. 

\paragraph{Comment about perfect cryptography.} 
The algebraic guard relation is based on the assumption that a term can only be derived algebraically, using the given operations and equations. A term $t$ thus either lies in a subalgebra generated by a set of terms $\Xi$, or not, and we have 
\bear
\Xi \vdash t & \vee & \Xi \not\vdash t
\eear
This means that the attacks on the implementation of the term $t$ are abstracted away. In particular, we assume that it is impossible to cryptanalyze the bitstrings representing $t$, and to derive $t$ by accumulating partial information about it. In other words, we assume {\em perfect cryptography}.

Moreover, we assume that the algebraic derivations $\Xi\vdash t$ only use the equations specified in the given algebraic theory $\TTt = (O,E)$. This means that the message algebra $\TTT$ is assumed to be a free $\TTt$-algebra, {\em or \/} that it is computationally unfeasible for the attacker to find any additional equations that $\TTT$ satisfies, not specified in the theory $\TTt$, and to use them in his derivations. This is roughly the {\em pseudo-free algebra\/} assumption \cite{RivestR:pseudo-free}.

\paragraph{Can we apply Thm.~\ref{CRprop} to the Hancke-Kuhn protocol?} The Hancke-Kuhn protocol on Fig.~\ref{HK} is obviously a timed version of the challenge response template from Fig.~\ref{CR}, for which Thm.~\ref{CRprop} provides a general security claim. If the guard condition holds, then the Theorem yields the security of the Hancke-Kuhn protocol. 

In the algebraic model, the attacker at a given state either knows a term, or not. As explained in Sec.~\ref{hancke}, the attacker on the Hancke-Kuhn protocol may always obtain half of the bits of the secret shared by Victor and Peggy by challenging her. Does this mean that the attacker gets to know the secret? If not, then the guard condition is satisfied. To apply Thm.~\ref{CRprop}, we should thus set up the algebraic model so that a term is known only when all of its bits are known.

Howeber, the same security proof would also hold for a modified version of the Hancke-Kuhn protocol, e.g. where $x\boxplus h = h^{(0)}$ if $x=a$ and $x\boxplus h = h^{(1)}$ otherwise, for some fixed $a\in \ZZz_2^\ell$. The attacker still cannot algebraically derive the term $x\boxplus h$ without $x$, because this term still depends on $x$. The guard condition holds, and thus the protocol is algebraically secure. In reality, though, the attacker who always responds with $h^{(1)}$ will succeed with a probability greater than $1-2^{-\ell}$, assuming that the challenge $x$ is drawn uniformly. The algebraic security of the Hancke-Kuhn type of protocols is not very realistic.

\section{Protocol models with guessing}\label{probguards}
In this section we propose a probabilistic refinement of the guard relation, which captures and quantifies just the partial information leaks, like the one in the Hancke-Kuhn protocol, without adding any unnecessary conceptual machinery.

\subsection{Implementing and guessing messages}
In order to reason about the feasibility of the algebraic operations on messages, and about guessing, we consider the {\em implementations\/} of the messages $t\in \TTT$ in an algebra $\Omega$ of {\em strings}, which carries the structure of a message $\TTt$-algebra, and moreover set of randomized functions. 

For concreteness, we assume that $\Omega = \ZZz^*_2$ is the set of bitstrings. However, any graded free monoid would do, since the only operations that we use are the concatenation and the length.

\subsubsection{Implementing messages}\label{Implementing messages}
\be{defn}
Let $H$ be a partially ordered set. We call an infinitely increasing chain $h_0\lt h_1\lt h_2\lt \cdots $ in $H$ a $H$-{\em tower}. We denote by $\sstr H$  the set of towers in $H$.
\ee{defn}

Any free monoid $\Omega$ is partially ordered by the {\em prefix\/} relation
\bear
a\sqsubset b & \iff & \exists c\in \Omega.\ a::c = b
\eear
where $a::c$ can be viewed as the concatenation of the strings $a$ and $c$. We call $\Omega$-towers {\em streams}. They are just infinite sequences of strings, strictly extending each other: a stream is a sequence $a = \{a_\ell\}_{\ell\in \NNn}\subseteq \Omega^\NNn$ such that $a_\ell \sqsubset a_{\ell+1}$ for all $\ell$. A stream $a$ is called an $\ell$-stream if the length of $\ell$-th element is exactly $|a_\ell|=\ell$. The set of streams through $\Omega$ is denoted by $\sstr\Omega$. 

$\NNn$ can be viewed as the special case, since a natural number can be viewed as  a string of 1s. The set $\sstr \NNn$ consists of strictly increasing sequences of natural numbers. 

\be{defn}
Let $\Var$ be a set of indeterminates. Its {\em strength\/} is a map $|-|:\Var\to \sstr \NNn$, assigning to each indeterminate $x$ for each value of the {\em security parameter\/} $\ell\in \NNn$ the required {\em length} $|x|_\ell\in \NNn$.  

An {\em environment} is a partial map $\eta:\Var \rightharpoonup \sstr\Omega$ such that $|\eta(x)_\ell| = |x|_\ell$ whenever $\eta(x)_\ell$ is defined.

An\/ {\em implementation\/} of a $\TTt$-algebra $\TTT$ is an injective $\TTt$-algebra homomorphism $\cod{-}\  :\  \TTT  \rightarrowtail \sstr\Omega$.

An environment and an implementation induce a $\TTt$-algebra homomorphism $\cod{-}_\eta :\TTT[\Var_\eta]\to \sstr\Omega$, where $\Var_\eta\subseteq \Var$ is the domain of definition of $\eta$. We call this homomorphism an implementation too whenever it is injective.
\ee{defn}

\paragraph{Explanations.} The string $\eta(x)_\ell \in\Omega$ is the implementation of the indeterminate $x$ with the security parameter $\ell$. The number $|x|_\ell$ is the required length of $x$ for the parameter $\ell$. The equation $|\eta(x)_\ell| = |x|_\ell$ enforces this requirement. Note that the function $|-|$ on the left is the length of the string in $\Omega$, whereas the function $|-|_\ell$ on the right is the part of the environment, specifying the required length.   

The implementation of the algebra $\TTT$ assigns a unique string to each term. By definition of the polynomial algebra $\TTT[\Var_\eta]$, every algebra homomorphism $\TTT\to \UUU$ to another algebra $\UUU$, and a function $\Var_\eta\to \UUU$ induce a unique algebra homomorphism $\TTT[\Var_\eta]\to \UUU$.  

We assume that any implementation is effectively invertible, i.e. that it is easy to recognize a term $t$ from its implementation $\cod t$.

Since any algebraic operation on $\Omega$ lifts to a pointwise operation over any power $\Omega^n$, it also lifts to streams. So $\sstr\Omega$ is also a $\TTt$-algebra, and a monoid for (elementwise) concatenation.\footnote{Grading is not an algebraic operation, and it does not lift: the length of each stream is infinite.}

\paragraph{Notation.} When confusion seems unlikely, we ignore the difference between the indeterminates $x,y\ldots \in \Var$ and their environment values $\eta(x), \eta(y)\ldots \in \Omega$.

\subsubsection{Randomized functions}
Consider the set of partial functions 
\begin{multline*}
\Randf  =  \{ \rf: \Omega \times \Omega \pfn \Omega\ |
 \forall x\forall \rho_1\forall \rho_2. \rf(\rho_1,a)\!\downarrow\ \wedge\ \rf(\rho_2,a)\!\downarrow\ \Rightarrow\ |\rho_1|=|\rho_2|\big\}
\end{multline*}
where $\rf(\rho,a)\!\downarrow$ means that $\rf$ is defined on $\rho, a$, and $\lvert \rho\rvert$ is the length of the bitstring $\rho$. The set ${\Randf}$ is a monoid with the following composition operation
\bear
\rf \circ \rg(\rho_2 :: \rho_1, a) & = &\rf(\rho_2, \rg(\rho_1,a)) 
\eear
and with the function $\iota \left(o, a\right) = a$ as the unit, where $o$ denotes the empty string. We interpret the elements of ${\Randf}$ as randomized functions over $\Omega$: the first argument $\rho$ represents the random seed, and the second argument $a$ is the actual input. The output $\rf a$ can then be viewed as a random variable with the probability distribution
\bea\label{Prob}
\Prob(\rf a = b) &  =&  \frac{\#\{\rho\ |\ \rf(\rho,a) = b\}}{2^r}
\eea
where $r$ is the length of all $\rho$ for which $\rf(\rho, a)$ is defined. Leaving the seed implicit, we denote randomized functions, as presented in ${\Randf}$, in the form $\rf:\Omega\rfun \Omega$.

\be{defn}
A {\em stream of functions\/} is a sequence $\rf=\{\rf_\ell\}_{\ell\in \NNn}\in {\Randf}^\NNn$ which is monotone, in the sense that for all streams $a,\rho \in \sstr \Omega$, at every $\ell\in \NNn$ holds
\begin{multline*}
\rf_\ell(\rho_\ell,a_\ell)\!\downarrow\ \, \wedge\ \ \rf_{\ell+1}(\rho_{\ell+1},a_{\ell+1})\!\downarrow \
\Longrightarrow\  \rf_\ell(\rho_\ell,a_\ell) \sqsubset \rf_{\ell+1}(\rho_{\ell+1},a_{\ell+1})
\end{multline*}
We denote the monoid 
of streams of functions  by $\sstr \Randf$.
\ee{defn}

\subsubsection{Indistinguishability}
\paragraph{Surviving the flood of negligible factors.} Every subterm of every term in every security protocol can in principle be guessed. Such probabilities are usually tolerably small: they are {\em negligible\/} functions of some security parameter $\ell$. In probabilistic analyses, it is often convenient to ignore such events of negligible probability. In a protocol analysis, tracking all terms and subterms that can be guessed with a negligible probability can lead to a lengthy list, without revealing anything non-negligible. In this section, we provide an underpinning for formal probabilistic reasoning {\em up to negligible factors}.

The frequencies of events are established by repeated sampling. The number of samples needed for a reasonable estimate depends on {\em a priori\/} chance that the event will occur. If this chance is 1 in $n$, then the number of the needed sample is an increasing function of $n$.

When sampling a stream $a = \{a_\ell\}_{\ell \in \NNn}$, we assume that a reasonable amount of samples should not be greater than $q(\ell)$, where $q$ is a function from a rig\footnote{A {\em rig\/} $Q$ is a "ring without the negatives": it consists of two commutative monoid structures, $(Q,+,0)$ and $(Q,\cdot,1)$,  such that $x\cdot(y+z) = x\cdot y + x\cdot z$ and $x\cdot 0 = 0$.} $Q\subseteq \NNn^\NNn$. In cryptography it is customary to take $Q = \NNn[x]$, the polynomials with non-negative integer coefficients. Streams are thus sampled a polynomial number of times. If the probability that the difference between $a_\ell$ and $b_\ell$ will be detected in $q(\ell)$ samples remains small for all $\ell$, then $a = \{a_\ell\}_{\ell \in \NNn}$ and $b = \{b_\ell\}_{\ell \in \NNn}$ are considered {\em indistinguishable}. In other words, $a$ and $b$ are indistinguishable if the probability that $a_\ell$ and $b_\ell$ are different is less than $\frac{1}{q(\ell)}$ for all $q\in Q$. Now we formalize this intuition. 

\be{defn}\label{negligible}
A function $\nu:\NNn\to [0,1]$ is said to be $Q$-{\em negligible\/} if it converges to $0$ faster than $\frac{1}{q(\ell)}$ for all $q\in Q$, i.e. 
\bear 
\forall q\in Q\ \exists n\in \NNn\ \forall \ell \geq n.\ \nu(\ell) & \lt &  \frac{1}{q(\ell)}
\eear
The set of $Q$-negligible functions is denoted by $\Qneg$.
The ordering on streams $a, b \in [0,1]^\NNn$ is defined up to negligible functions, i.e.
\bear
a\leq b & \iff & \exists \nu \forall \ell.\ a_\ell +\nu(\ell) \leq b_\ell
\eear
We say that $a, b \in [0,1]^\NNn$ are $Q$-{\em indistinguishable}, and write $a \stackrel Q \sim b$, if $a\leq b$ and $b\leq a$, or equivalently
\bear
a\sim b & \iff & \exists \nu \forall \ell.\ |a_\ell - b_\ell|  \leq \nu(\ell)
\eear
\ee{defn}

\paragraph{Assumption, examples.} For simplicity, we take $Q$  to be the rig  $\NNn[x]$ of polynomials with non-negative integer coefficients, as it is usually taken in cryptography. Then, e.g., for $a = \{2^\ell\}_{\ell\in \NNn}$ and $b = \{\ell^{-2}\}_{\ell \in \NNn}$ holds $a \sim 0$, but $b \not \sim 0$, where $0$ is viewed as the constrant sequence.

\be{defn}
Streams of functions $\rf$ and $\rg$ are  indistinguishable if the sequences $\Prob(\rf a=b)$ and $\Prob(\rg a = b)$ are indistinguishable for all streams $a,b\in \sstr \Omega$. We abbreviate
\bear
\rf\sim \rg & \iff & \forall ab\in \sstr \Omega.\ \Prob(\rf a=b) \sim \Prob(\rg a = b)
\eear
\ee{defn}

\be{defn}
A\/ {\em flow}
is an equivalence class of streams of randomized functions. The flow monoid $\Rand$ is thus
\bear
\Rand & = & \sstr\Randf / \sim
\eear
\ee{defn}

\subsection{Probabilistic derivability}
In contrast with the algebraic derivability relation from Sec.~\ref{States}, the probabilistic derivability relation does capture partial information leaks, using the implementations of the terms. While $\Xi \not \vdash \Theta$ may happen because some $t\in \Theta$ is not algebraically derivable from $\Xi$, it may be easy to guess many bits of information about $\Theta$ from $\Xi$. We formalize this by saying that for some stream of randomized functions $\rf\in {\Randf}$, $\Prob(\rf\cod \Xi = \cod\Theta)$ is high. By assumption, the messages $\Theta$ are easily decoded from their implementations $\cod \Theta$. So if some $\rf$ is likely to output $\cod \Theta$ on the input $\cod \Xi$, then the chance to derive $\Theta$ from $\Xi$ is high. This is what we want to capture by the following {\em randomized\/} derivability relation, which quantifies {\em guessing chance}.

Let $\FV\Xi\subseteq \Var$ be the set of indeterminates that occur in $\Xi$. Any minimal environment $\eta$ in which the $\cod \Xi_\eta$ is defined must be defined over $\FV \Xi$. Since for each $\ell$ the required number of bits for each $x\in \FV\Xi$ is fixed to $|x|_\ell$, each $\eta_\ell$ must select the same number of bits 
\bear
|\FV \Xi |_\ell & = & \sum_{x\in \FV\Xi} |x|_\ell
\eear
So there are $2^{|\FV \Xi|_\ell}$ environments to interpret $\Xi$ for the security parameter $\ell$.
Our chance to guess $\Theta$ from $\Xi$ is the probability that a flow $f\in \Rand$ will output $\cod \Theta_{\eta}$ when given the input $\cod \Xi_{\eta}$, for the random choices of $\eta$. Hence the following definition.

\be{defn}
The {\em guessing chance\/} $\pder \Xi \Theta$ is the stream of probabilities 
\bea\label{probder} 
\pder \Xi \Theta_\ell & =& \bigvee_{\rf_\ell \in {\Randf}}  \frac{\#\{\eta_\ell\ |\ f_\ell\cod \Xi_\ell = \cod\Theta_\ell\}}{2^{|\FV{\Xi,\Theta}|_\ell}}
\eea
viewed up to indistinguishability.

We abbreviate $\pder \emptyset \Theta$ to $\ppder \Theta$.
\end{defn}

\paragraph{Comment.} It is assumed that the guesser evolves: he keeps finding better and better randomized functions $f_\ell$, and thus computes the supremum in \eqref{probder} as the time goes by\footnote{This kind of spontaneous optimization underlies dynamics of evolutionary processes in general \cite{Maynard-Smith}.}. 

Since the functions in the sequence $\{f_\ell\}_{\ell\in \NNn}$ compute on streams $\cod \Xi_\ell$, together they form a stream of functions $f\in \sstr\Randf$, i.e. a flow $f\cod \Xi = \Theta$.

\paragraph{Examples.}  For any closed term $t\in \TTT$, i.e. such that $\FV t = \emptyset$, it holds that $\ppder t = 1$. To see this, note that $\cod t$ is given in the empty environment $\eta_\emptyset$, and thus $\FV t = \emptyset$ implies $|\FV t|_\ell = 0$ for all $\ell$. 
By the assumption about $\Randf$, for every stream $\cod t\in \sstr\Omega$, the constant function stream $f() = \cod t$ is feasible. 
The supremum of \eqref{probder} is reached at the constant function stream $f() = \cod t$, and gives $\ppder t = \frac{\#\{\eta_\emptyset\ |\ f()=\cod t\}}{2^0} = 1$.

On the other hand, for every $x\in \Var$ holds  $\ppder x_\ell = 0$. There are exactly $2^{|x|_\ell}$ environments $\eta_x$, defined on $x$ alone. To guess $x$ without any inputs, we need a constant flow $f$, such that $f()=\cod x = \eta_x (x)$, i.e. a constant stream of functions $f_\ell() = \eta_x(x)_\ell$. Whichever $f$ we may choose, exactly one environment $\eta_x$ will give $f()=\eta_x (x)$. So for every constant flow $f$ holds $\frac{\#\{\eta_x\ |\ f()_\ell = \cod x_\ell\}}{2^{|x|_\ell}} = \frac{1}{2^{|x|_\ell}}$. The supremum in \eqref{probder} is thus reached for all constant $f\in \Rand$, and $\ppder x_\ell = \frac{1}{2^{|x|_\ell}}$. But the sequence $\left\{{2^{-|x|_\ell}}\right\}_{\ell\in \NNn}$ is indistinguishable from $0$, as pointed out after Def.~\ref{negligible}.

\subsubsection{Subbayesian reasoning and Advantage}
\be{prop}\label{pseudobayes}  For all sets of terms $\Xi,\Gamma,\Theta$ holds
\bea
\pder{\Xi}{\Gamma} \cdot \pder{\Xi,\Gamma}{\Theta} & \leq & \pder{\Xi}{\Gamma,\Theta} \label{comp}
\eea
When $\ppder\Gamma \gt 0$, it follows that 
\bea \label{fractn}
\pder{\Gamma} \Theta & \leq & \frac{\ppder {\Gamma,\Theta}}{\ppder{\Gamma}} \label{MP} 
\eea
\ee{prop}
The inequalities become equalities if $\Xi$ and $\Theta$ have no indeterminates in common.

\subsubsection{Advantage}
\be{defn}\label{advantage} The\/ {\em advantage} provided by a set of terms $\Xi$ in computing the terms $\Theta$ is the value
\bear
\adv \Xi \Theta & = & \pder \Xi \Theta - \ppder \Theta
\eear
When this advantage is zero, we say that $\Theta$ is\/ {\em  flow independent} of $\Xi$, and write 
\bear
\indep \Xi \Theta  \iff  \adv \Xi \Theta = 0
 \iff  \pder \Xi\Theta = \ppder\Theta 
\eear
\ee{defn}

\subsection{Probabilistic guards}
The idea of the guard relation is that a term $t$ is guarded by one of the guards from $\GGG$ if whenever $t$ is derived, then at least one of the guards $\Gamma\in \GGG$ is also derived. In the algebraic model, this was simple enough to state by Definition \ref{alguards}. When $t$ can be guessed, then this crude statement needs to be refined: the event that $t$ is guessed must be preceded by the event that some $\Gamma\in \GGG$ is guessed. 

\be{defn}\label{compguards}
We say that a set of sets of terms $\GGG$\/ {\em guards (against guessing)} a term $t$ with respect to a set of terms
$\Upsilon$, and write ${\GGG}\ \mbox{{\sf \small guards}}\  {t}\ \mbox{{\sf \small within}}\ {\Upsilon}$ if for all $\Xi\subseteq \Upsilon$ such that $\adv{\Xi}{t} \gt 0$ holds 
\bea\label{compguardseq}
\pder \Xi t   & \leq &    \bigvee_{\Gamma\in \GGG} \pder \Xi \Gamma \cdot \pder{\Xi,\Gamma}{t}
\eea
\ee{defn}

\paragraph{Explanation.} In the algebraic case, \eqref{alguardseq}  was an attempt to capture the intuition that $\GGG$ guards $t$ if all computational paths to $t$ lead through some $\Gamma\in \GGG$, assuming the context $\CCC$. The above definition extends this attempt to computational paths {\em with guessing}.  If we get any help from $\Xi$ to guess $t$, then that help is not greater than the help we get from it to guess some guard $\Gamma \in \GGG$ of $t$ first, and then to guess $t$ from this guard. Applied to message theories with trivial implementations (e.g. with $\Omega = 1$), Def.~\ref{compguards} boils down to Def.~\ref{alguards}, in the sense that the guessing chance is always constantly 0 or constantly 1, and \eqref{compguardseq} reduces to \eqref{alguardseq}.
 
 \be{prop}\label{probIsDet}
Suppose that the guessing machines ${\Feasf}$ used in \eqref{probder} are constrained to never read their random bits, so that guessing boils down to algebraic derivations. Then the guessing guard relation from (\ref{compguardseq}) boils down to the algebraic guard relation from (\ref{alguardseq}).
 \ee{prop}

To simplify notation, we elide the environment subscripts from $\cod{-}_\eta$ whenever $\eta$ is inessential for the argument.

\section{Partitioned functions and $\boxplus$} \label{boxplus}
\paragraph{Notation.} Whenever the confusion is unlikely, we abuse notation and denote by $x$ both the indeterminate $x\in \Var$ and its implementation $\cod x = \eta(x) \in \Omega$.

In this section we analyze a class of quickly computable functions, like the one used in the Hancke-Kuhn protocol. One way to ensure that a function is  quickly computable is to require that the bit dependency of its outputs from its inputs must be partitioned: the $i$-th block of output bits should only depend on the $i$-th block of input bits. Obviously, a function where every bit of output depends on every bit of input has to wait for the last bit of input before it can produce.
Since in this section we are dealing with purely random input, our results are presented in terms of streams, not flows.


\begin{defn}
We say that a boolean function $f:\ZZz_2^m \to \ZZz_2^n$ is {\em partitioned\/} when 
\bear
m & = & m_1+m_2 +\cdots + m_{\ell}\\
n & = & n_1 + n_2 + \cdots + n_{\ell}\\
f & = & f_1\ ::\ f_2\ ::\ \cdots\ ::\ f_{\ell}
\eear
where   $f_i:\ZZz_2^{m_i} \to \ZZz_2^{n_i}$, for $i = 1,2,\ldots \ell$ are independent on the inputs and the outputs of all other component functions, in the sense that $\indep{x_{\overline \imath}, f_{\overline \imath}(x_{\overline \imath})}{f_i(x_i)}$, where $\overline \imath = \{j\leq \ell|\  j\neq i\}$.
\ee{defn}

Clearly, a boolean function receiving its input string sequentially can already return the $i$-th block of its outputs while still receiving $i+1$st block of the inputs. Unfortunately, this convenient property also decreases cryptographic strength of the function, which requires that each bit of the output depends on each bit of the input \cite{Webster-Tavares}.
In particular, knowing a value $f(z)$ of a partitioned function increases the chance of guessing $f(x)$. We make this precise in the next section.

\subsection{Guessing partitioned functions}

\be{prop}\label{fzfx}{\bfseries (a)} Let $f$ be a randomized partitioned function, and let $x,z\in \ZZz_2^m$ be fixed bitstrings with a common block $x_i = z_i\in \ZZz_2^n$. Then $\pder {x,z,f(z)}{f(x)} \geq 2^{n - m}$.

\noindent{\bfseries (b)} Let $f:\ZZz_2^\ell\to\ZZz_2^\ell$ be randomized {\em bitwise} partitioned, i.e. $\lvert m_i\rvert =\lvert n_i\rvert = 1$
 for all $i\leq \ell$. Then  $\pder { x,z,f(z)}{f(x)} \geq 2^{-\Delta(x,z)}$, where $\Delta(x,z) = \#\{i | x\neq z\}$ is the Hamming distance.
\ee{prop}


A consequence of Prop.~\ref{fzfx} is that a proximity authentication protocol, implemented using a partitioned function $R$ to compute the response $r^{VP} x = R(s^{VP},c^{VP}x)$, cannot be secure in an absolute sense, because the response may  be guessed with a non-negligible probability from the other responses $r^{VP} z$. Moreover, it seems that the attacker can always obtain some other responses $r^{VP} z$ by impersonating Victor and  issuing challenges $c^{VP} z$.

\begin{lemma}\label{bitwise}
A randomized boolean function $f:\ZZz_2^\ell \to \ZZz_2^\ell$ is bitwise partitioned if and only if for every $x\in \ZZz_2^\ell$ it holds that
\bea
f(x) &  = & x\boxplus \left(f(0^\ell)\ ::\ f(1^\ell)\right)
\eea
where $\boxplus$ is the Hancke-Kuhn function (\ref{HKfun}), and $0^\ell, 1^\ell\in \ZZz_2^\ell$ are the strings of 0s and 1s, respectively.
\end{lemma}

Bitwise partitioned functions with a minimal guessing probability can now be completely characterized: they turn out to be {\em precisely\/} the Hancke-Kuhn functions (\ref{HKfun}) for which the values at 0 and at 1 are independent.

\be{prop}\label{eqzx} Suppose that  $f:\ZZz^\ell_2 \to \ZZz_2^\ell$ is a randomized bitwise partitioned function such that
$\indep{x}{f(0^\ell) :: f(1^\ell)}$. Then  for fixed $z$ and $x \in \ZZz^\ell$: 
\bea\label{crude}
\pder {x, z,f(z)}{f(x)} & = & 2^{- \Delta(z,x)}
\eea
if and only if for every $i\leq \ell$ it holds that 
\bea\label{bxpl}
\indep{f_i(0)}{f_i(1)}&\mbox{and} & \indep{f_i(1)}{f_i(0)}
\eea
\ee{prop}

\paragraph{Remark.} In a sense, $x\boxplus (-):\ZZz_2^{2\ell}\to \ZZz_2^\ell$ is thus a "one-and-half-way function", since $x\boxplus h$ discloses only one half of the bits of $h$. 

On the other hand, $(-)\boxplus h: \ZZz_2^\ell \to \ZZz_2^\ell$ is not only an example of a bitwise partitioned function, satisfying the needs of the Hancke-Kuhn protocol, but it is a canonical way to represent such functions.  

\subsection{Guessing $x\boxplus h$}
We now consider the probability of guessing $x \boxplus h$ given various sorts of information that may be
learned in the Hancke-Kuhn protocol.

\be{defn}\label{kerdef}
a) For $x\in \ZZz_2^\ell$ and $I\subseteq \ell = \{0,1,2,\ldots \ell-1\}$ we define $x^{\wcard I}\in \ZZz_2^\ell$ to be the bit string obtained by replacing for all $i\in I$ the bits $x_i$ with a ``wild card'' $\wcard$
\bear
x^{\wcard I}_j & = & \begin{cases}
\wcard & \mbox{ if } j\in I\\
x_j & \mbox{ otherwise}
\end{cases}
\eear
b) For $h = h^{(0)}::h^{(1)}$, where $h^{(0)}, h^{(1)} \in \ZZz_2^{\ell}$ we define the\/ {\em kernel} $\kernel h$ to be the set of places where its first and its second half coincide, e.g.
\bear
\kernel h & = & \{i \in \ell\ |\ h^{(0)}_{i} = h^{(1)}_{i}\}.
\eear
\ee{defn}
We make use of these definitions in the following.

\be{prop}\label{kernel}
Suppose that $h$ the concatenation of two constant $\ell$-bit streams, and $x$ is a uniformly distributed $\ell$-bit stream.  Then
\begin{anumerate}
\item $\pder{h}{x \boxplus h}_\ell = 2^{| \kernel h | - \ell}$
\item $\pder{x, h}{x \boxplus h}_\ell = \pder{\subs x {\kernel h}, h }{x \boxplus h}$
\end{anumerate}
\ee{prop}

The following lemma concerns the problem of deriving $x \boxplus h$ from $z \boxplus h$ for some $z$.

\be{prop}\label{monty}
Let $h$ be the concatenation of two uniformly distributed $\ell$-bit streams, let $x$ be a uniformly distributed $\ell$-bit stream, and let $z$ be any $\ell$-bit stream.  Then the following holds.
\bear
\pder{ z \boxplus h}{x \boxplus h}_\ell & = & \pder{z,  z \boxplus h}{x \boxplus h}_\ell \  =\   \left(\frac{3}{4}\right)^{\ell}
\eear
 \ee{prop}

\section{Security of Hancke-Kuhn}
\label{proof}

We quantify the security of the Hancke-Kuhn protocol by evaluating $\Prob({\sf crp})$, i.e. the probability that the sequence of events in a complete protocol run validates the following reasoning of Victor's
\begin{multline}
V : \ \ (\nu x)_V \rar \startt<x>_V \rar  \stopp(x\boxplus \token)_{V} \notag\\ 
\Longrightarrow \ \Big((\nu x)_V \rar \startt<x>_V \rar (x)_P \rar <x\boxplus \token>_{\orig P}\rar
\stopp(x\boxplus \token)_{V}\Big) \tagg{crp} 
\end{multline}
corresponding to the run on Fig.~\ref{HK}.
In order to evaluate this probability, we analyze the probability that {\sf (crp)} fails. How can it happen that Victor observes a satisfactory sequence of his own actions 
\bea \label{vvv}
\VVV & = & (\nu x)_V \rar {\startt}<x>_V\rar{\stopp}\left(x\boxplus \token\right)_{V}
\eea
but that the desired run
\bea \label{ooo}
\OOO \  = \  {\startt}<x>_V\rar(x)_P \rar 
\left<x\boxplus \token\right>_{\orig P}\rar {\stopp}\left(x\boxplus \token\right)_{V}
\eea
did not take place? There are just two possibilities:
\begin{description}
\item[$\AAA$:] the responder does not know the secret $s$, i.e. he is the $\AAA$ttacker,
\item[$\EEE$:] the responder knows the secret $s$, i.e. he is Peggy, but the response is sent $\EEE$arly, without receiving the challenge.
\end{description}
The remaining case, that the responder is Peggy, and she responds to the challenge, is just the event $\OOO$. Thus $\neg \OOO = \AAA\cup \EEE$. It follows that
\bea
\Prob({\sf crp})\ &  = &\ \Prob(\OOO| \VVV)\ =\ 1-\Prob(\AAA \cup \EEE| \VVV)\notag \\ 
& \geq &\ 1-\Prob(\AAA| \VVV) - \Prob(\EEE| \VVV) \label{un}
\eea
The (in)security of the Hancke-Kuhn protocol thus boils down to evaluating $\Prob(\AAA|\VVV)$ and $\Prob(\EEE|\VVV)$. The following lemmas and propositions show that these probabilities are negligible. The  proofs are in the Appendix.

\paragraph{Response token.}
Recall that Peggy's response token $\token = H(s::a::b)$ is derived from  the shared secret $s$, Peggy's counter $a$, and Victor's counter $b$, using a secure public hash function $H$. In this section, ${\token}$ abbreviates $H(s::a::b)$. 

\be{assm}\label{assumption} {\rm The above decomposition of $\neg \OOO$ as $\AAA\cup \EEE$ is valid only if $\token = H(s::a::b)$ is such that
\begin{itemize}
\item $|s|\gg|x|$, i.e. attacker's chance to guess the secret $s$ is negligible compared with his chance to guess the challenge $x$;
\item the counters $a$ and $b$ are never reused (although they may be predictable).
\end{itemize}
Otherwise, the attacker may guess $\token$, and $\neg \OOO$ may not be covered by $\AAA\cup \EEE$.}
\ee{assm}

\subsection{Guards in undesired runs}
In order to evaluate $\Prob({\sf crp})$, we need to determine the probability that the correct response $x\boxplus {\token}$ is guessed in the undesired runs $\AAA$ and $\EEE$. Towards this goal, we explore
 what can be guessed in the term contexts (cf. Def.~\ref{termcont})  $\AAA(x\boxplus {\token})$ and $\EEE(x)$. The following lemmas simplify this question.

\begin{lemma}\label{sets} 
\begin{cond}
Let $\AAA$ be an attack run with a long term secret $s$, Peggy's counter $a$, Victor's counter $b$, and $\AAA$ttacker's challenge $z$, for which he obtains the response $z\boxplus {\token}$, where ${\token} = H(s::a::b)$. Then for any $\Xi\subseteq \crbefo{\AAA}{x \boxplus {\token}}$ it holds that
\bear
 \pder{\Xi}{x \boxplus {\token}} & = &  
 \pder{\Xi \cap \left\{s, a, b,x,z, z\boxplus {\token}\right\}}{x \boxplus {\token}}
\eear
 \end{cond}
\be{cond} Let $\EEE$ be a run with a long term secret $s$, Peggy's counter $a$, Victor's counter $b$, and where Peggy responds early. Then for any $\Xi\subseteq\crbefo{\EEE}{x}$ it holds that
\bear
 \pder{\Xi}{x \boxplus {\token}}  & = &  
 \pder{\Xi \cap \left\{s, a, b \right\}}{x \boxplus {\token}}
 \eear
\end{cond}
\end{lemma}

\begin{lemma}\label{grindout}
For $h = H(s::a::b)$ and $\Upsilon \subseteq \left\{z,  z\boxplus {\token} \right\}$ it holds that
\begin{gather}
\ppder{x \boxplus {\token}}_{\ell}= \pder{x,z}{x\boxplus \token}_{\ell} = 2^{-\ell}\label{aaa}\\ 
\pder{a,b,s,\subs x {\kappa {\token}}}{x \boxplus {\token}} = 1 \label{bbb}\\
\pder{a,b,s,x,\Upsilon}{x \boxplus {\token}} = 1 \label{ccc}
\end{gather}  
\end{lemma}

\begin{prop}\label{guardprop}
$\left\{ \{s\},  \{   z \boxplus {\token}\} \right\}  \mbox{ \sf guards } x \boxplus {\token} 
\mbox{ \sf within } \crbefo{\AAA}{x \boxplus {\token}}$
\end{prop}

 \begin{prop}\label{guardprop2}
$\left\{ \{\subs x {\kappa {\token}}\} \right\} \mbox{ \sf guards } x \boxplus {\token} \mbox{ \sf within } \crbefo{\EEE}{x} $
\end{prop}

The guards displayed in the preceding Propositions will now be used to evaluate $\Prob(\VVV|\AAA)$ and $\Prob(\VVV|\EEE)$, i.e.  the probabilities that the authentication may fail because the $\AAA$ttacker breaks it, or because Peggy's succeeds in responding $\EEE$arly.

\subsection{Bounds on undesired runs}
Proposition \ref{guardprop} and the definition of probabilistic guards say that, for a given challenge $x$, the probability that an $\AAA$ttacker can violate authentication is bounded above by 
\begin{gather*}
\pder{\Phi}{s} \cdot \pder{\Phi, s}{x \boxplus {\token}}  \mbox{ or by } \\
\pder{\Phi}{z \boxplus {\token}} \cdot \pder{\Phi, z \boxplus  {\token}}{x \boxplus {\token}}
\end{gather*}
\noindent where $\Phi = \{ a,  b, z, z \boxplus {\token} \}$.  The first quantity is clearly negligible. We must show the same for the second.

Likewise, Proposition \ref{guardprop2} implies that the probability that Peggy can respond $\EEE$arly is bounded above by
\[ \pder{s,a,b}{\subs{x}{\kernel {\token}}} \cdot \pder{ s,a,b, \subs{x}{\kernel {\token}}}{x \boxplus {\token}}
\]
Note that in the attack run $\AAA$, the $\AAA$ttacker cannot learn $x$ until after she has created $z$.  The distribution of $z$ is thus independent from that of $x$.

\be{prop}\label{34}
Suppose that the $\AAA$ttacker, before receiving Victor's challenge $x$, can pick her own challenge $z$ and obtain a single response $z \boxplus {\token}$. Then the stream of expected probabilities  $\Prob(\VVV |\AAA)$ 
that the $\AAA$ttacker can deceive Victor by guessing  $x \boxplus {\token}$ is indistinguishable from the stream of probabilities $p$ defined by
\bear
p_\ell = \sum_{x\in \ZZz_2^\ell} 2^{-\ell}\pder{x,z,z\boxplus {\token}}{x\boxplus {\token}}_\ell &= &  \left(\frac{3}{4}\right)^{\ell}
\eear
This means that $\Prob(\VVV |\AAA)$  is negligible.
\ee{prop}

\be{prop}\label{early}
The stream of expected probabilities $\Prob(\VVV |\EEE)$ that Peggy can deceive Victor by guessing and sending her response before she receives the challenge  is indistinguishable from the stream $q$ defined by 
\bear
q_\ell  = \sum_{\token \in \ZZz_2^{\ell}} \sum_{x \in \ZZz_2^{\ell}}2^{-\ell} \pder{\token}{x \boxplus \token}_\ell & = & \left(\frac{3}{4}\right)^\ell
\eear
This means that $\Prob(\VVV |\EEE)$  is negligible.
\ee{prop}

Note in particular that this means that in both cases the stream of probabilities is  indistinguishable from zero, since the stream  $\left(\frac{3}{4}\right)^{\ell}$ is itself indistinguishable from zero.

The final result is obtained by putting Propositions \ref{guardprop} and \ref{34} together.  

\be{thm}\label{mainthm}
Suppose that the Hancke-Kuhn protocol is realized in such a way that it satisfyes \ref{assumption}, and does not always fail for trivial reasons: i.e., there are some sessions with an honest prover Peggy and an honest verifier Victor. Formally, this means that there are $C,D \in (0,1)$ such that
\begin{itemize}
\item $\Prob(\AAA), \Prob(\EEE)\lt C$, i.e. not every response is from an $\AAA$ttacker, or too $\EEE$arly,
\item $\Prob(\VVV)\gt D$, i.e. Victor sometimes observes a satisfactory run and accepts.
\end{itemize}
Then $\Prob{\sf (crp)}$ is indistinguishable from 1. In other words, the Hancke-Kuhn protocol achieves authentication almost certainly. 
\ee{thm}

\section{Conclusion}\label{conclusion}
We have presented a framework for extending algebraic cryptographic models to probabilistic models and used it to construct a probabilistic extension of the Protocol Derivation Logic.  We have illustrated it by applying it to an analysis of the Hancke-Kuhn distance bounding protocol.   We expect that it will be useful in the analysis of many other protocols that
rely on weak cryptography to take advantage of non-standard communication channels.

We should also point out that the potential applications of our framework go far beyond purely probabilistic extensions.  The main thing that needs to be done to make our framework applicable to computational models is to define a notion of feasibly computable functions, so that guessing probability can be defined in terms of feasible function streams instead of all possible function streams.  We have defined such a notion and are currently investigating its applications to protocols.  In future work, we expect to present a more general framework that can incorporate a wide range of methods of cryptographic reasoning.

\paragraph{Acknowledgement.} We are grateful to Joshua Guttman, John Mitchell, Mike Mislove, and to several anonymous referees for careful reading of earlier versions of this paper, and for valuable suggestions towards improvements in presentation.

\bibliography{ref-proximity,ref-perdy}
\bibliographystyle{plain}

\appendix
\section{Appendix: The Proofs}
\bprf{ of Prop.~\ref{pseudobayes}} Let $f_\ell$ and $g_\ell$ be randomized functions. Consider the sets $F = \{\chi_{\ell}\ |\ f_\ell\cod \Xi_{\chi\ell} = \cod\Gamma_{\chi\ell}\}$ and $G = \{\eta_{\ell}\ |\ g_\ell\cod {\Xi,\Gamma}_{\eta\ell} = \cod\Theta_{\eta\ell}\}$.
\begin{description}
\item[Claim 1.] If for $x,y \in\FV{\Xi,\Gamma}$ and $\eta_\ell$ such that $g_\ell\cod {\Xi,\Gamma}_{\eta\ell} = \cod\Theta_{\eta\ell}$ holds $\eta_\ell(x) = \eta_\ell(y)$, then for $\widehat\eta_\ell$, which is equal to $\eta_\ell$ everywhere except on $\widehat\eta_\ell(x) \neq \widehat\eta_\ell(y)$, holds that  $\widehat g_\ell\cod {\Xi,\Gamma}_{\widehat\eta\ell} = \cod\Theta_{\widehat\eta\ell}$, for $\widehat g$ modified accordingly. (Intuitively, separating two pieces of input can only provide more information, not less.)
\item[Claim 2.] If $f_\ell\cod \Xi_{\chi\ell} = \cod\Gamma_{\chi\ell}\}$ and $\dom(\chi_\ell) \subseteq \dom(\eta_\ell)$, with $\chi_\ell(x) \neq \chi_\ell(y)\Rightarrow \eta_\ell(x) \neq \eta_\ell(y)$, then $f_\ell$ can be precomposed with a permutation to yield $\widehat f_\ell$ with $\dom(\widehat f_\ell)\subseteq \dom(\eta_\ell)$ and $\widehat f_\ell\cod \Xi_{\eta\ell} = \cod\Gamma_{\eta\ell}\}$.  
\end{description}
The consequence of these claims is that we can modify $f_\ell$ and $g_\ell$ to $\widehat f_\ell$ and $\widehat g_\ell$ so that $\#F = \#\widehat F$ and $\# = \widehat G$.

Now let $h_\ell(x) = f_\ell (x)::g_\ell (x::y)$. Since thus 
\[
h_\ell \cod\Xi_{\eta\ell} \ = \ 
\left(f\cod\Xi_{\eta\ell}\right):: \left(g\left(\cod{\Xi}_{\eta\ell} :: f\cod\Xi_{\eta\ell}\right)\right)
\ =\ 
\cod{\Gamma,\Theta}_{\eta\ell}
\]
holds, we have
\begin{multline*}
\frac{\#\{\eta_{\ell}\ |\ f_\ell\cod \Xi_{\ell} = \cod\Gamma_{\ell}\}}{2^{|\Xi,\Gamma,\Theta|_{\ell}}} \ \cdot \ \frac{\#\{\eta_{\ell}\ |\ g_\ell\cod {\Xi,\Gamma}_{\ell} = \cod\Theta_{\ell}\}}{2^{|\Xi,\Gamma,\Theta|_{\ell}}} \ \leq \\
 \frac{\#\{\eta_{\ell}\ |\ h_\ell\cod \Xi_{\ell} = \cod{\Gamma,\Theta}_{\ell}\}}{2^{|\Xi,\Gamma,\Theta|_{\ell}}}
\end{multline*}
The inequality $\pder{\Xi}{\Gamma} \cdot \pder{\Xi,\Gamma}{\Theta} \leq \pder{\Xi}{\Gamma,\Theta}$ follows by observing that
\bear
\frac{\#\{\eta_{\ell}\ |\ f_\ell\cod \Xi_{\ell} = \cod\Gamma_{\ell}\}}{2^{|\Xi,\Gamma,\Theta|_{\ell}}} & = & \frac{\#\{\chi_{\ell}\ |\ f_\ell\cod \Xi_{\ell} = \cod\Gamma_{\ell}\}}{2^{|\Xi,\Gamma,|_{\ell}}}
\eear 
\epr


\bprf{ of Prop.~\ref{fzfx}} 
For (a), $x_i = z_i$ yields $f_i(x_i)=f_i(z_i)$, so we only need to guess at most $n-n_i$ bits. For (b), $x_i$ and $z_i$ are bits, and $n-\Delta(x,z)$ of them are equal, so we only need to guess at most $\Delta(x,z)$ bits.
\epr

\bprf{ of Lemma \ref{bitwise}}
$\left(f(x)\right)_i \  = \  f_i(x_i)\ =\ 
\left(x\boxplus \left(f(0^\ell)\ ::\ f(1^\ell)\right)\right)_i$ holds by the definition of bitwise partitioned functions at the first step, and by (\ref{HKfun}) at the second step.
\epr

\bprf{ of Prop.~\ref{eqzx}}
Assumptions (\ref{bxpl}) say that 
\bear
x_i\neq z_i \ &\Longrightarrow & \ \pder{x_i,z_i,f_i(z_i)}{f_i(x_i)} = \pder{x_i}{f_i(x_i)}
\eear 
On the other hand, by definition, the components of a partitioned function are mutually independent. 
Hence 
\bear
\pder{x,z,  f(z)  }{f(x)}\ & = &\ \prod_{i=1}^\ell \pder{x,z,f(z) }{f_i(x_i)}  \ =\  \prod_{\substack{i=1\\x_i\neq z_i}}^{\ell} \pder{x_i}{f_i(x_i)} \\ 
 &=& \ \prod_{\Delta(z,x)}\frac{1}{2}\ = \ 2^{-\Delta(z,x)} 
\eear

The other way around, using (\ref{crude}) at the second step, we get 
\bear
\prod_{i=1}^\ell \pder{x,z,f(z) }{f_i(x_i)} & = &  \pder {x, z,f(z) }{f(x)}  
\ =\     2^{- \Delta(z,x)} \\
& = &  \prod_{\substack{i=1\\x_i\neq z_i}}^{\ell} \pder{x_i}{f_i(x_i)}
\eear
which, with the componentwise independence, yields (\ref{bxpl}).
\epr

\bprf{ of Prop.~\ref{kernel}}
Note that for each $i\in \kernel h$, the bit $(x\boxplus h)_i = h^{(0)}_i = h^{(1)}_i$ does not depend on $x_i$. This means that $x\boxplus h$ only depends on $\subs x {\kernel h}$.
\epr

\bprf{ of Prop.~\ref{monty}}
Guessing $x\boxplus h$ from $z$ and $z\boxplus h$ can be modeled as a version of the Monty Hall problem \cite{monty-hall}, where Monty randomly selects $x$ and $h$ and the contestant chooses $z$. Monty then announces $z\boxplus h$ and the contestant guesses $x\boxplus h$. 

Since the bits of $x\boxplus$ are independent, it is enough to consider the case $\ell = 1$. Monty then flips three fair coins to pick the secret bits $x, h^{(0)}$, and $h^{(1)}$, while the contestant picks a bit $z$.   Monty then announces $z\boxplus h = h^{(z)}$. Should the contestant now guess that $x\boxplus h = z\boxplus h$, or should he switch to $x\boxplus h = \neg (z\boxplus h)$?

Denote by $q$ the probability that the contestant picks $x\boxplus h = z\boxplus h$. If $h^{(0)} = h^{(1)}$, the contestant wins with this choice, because the value $x\boxplus h$ is the same for every $x$. Since $h^{(0)}$ and $h^{(1)}$ were randomly chosen, $\Prob(h^{(0)} = h^{(1)}) = \frac{1}{2}$. Otherwise, if  $h^{(0)} \neq h^{(1)}$, then $x\boxplus h = z\boxplus h$ holds if and only if $x=z$. Since $x$ is random, $\Prob(x=z) = \frac{1}{2}$, and hence $\Prob(h^{(0)} \neq h^{(1)} \wedge x = z) = \frac{1}{4}$, because $h^{(0)}$, $h^{(1)}$ and $x$ are independent. 

The probability that the contestant will make a correct guess is thus 
\bear
q\cdot \left( \Prob\left(h^{(0)} = h^{(1)}\right)+\Prob\left(h^{(0)} \neq h^{(1)} \wedge x = z\right)\right)
=\ \ \frac{3q}{4}
\eear
To maximize this probability, the contestant needs $q=1$, and should thus stick\footnote{This solution is in contrast from the original Monty Hall problem \cite{monty-hall}, where it is advantageous to switch. The reasoning is, however, quite similar.} 
with Monty's bit $z\boxplus h$.

The proof for $\pder{z \boxplus h}{x \boxplus h}$ differs just in the detail that $z$ is not chosen by the contestant, but obeys some unknown distribution. However,
$x$ is still independent of $z$. 
Thus for some $p$, $\Prob(x=z) = Prob(x=0)\cdot Prob(z=0)+Prob(x=1)\cdot Prob(z=1) = \frac{1}{2}p+\frac{1}{2}(1-p)=\frac{1}{2}$.
\epr

\bprf{ of Lemma \ref{sets}{\em (a)}}
By assumption, the outputs of the hash function $H$ are indistinguishable from random strings, and thus satisfy $\indep{H(u)}{H(v)}$ for all $u\neq v$.

Recall that $\crbefo{\AAA}{x \boxplus {\token}}$ is the union of the contexts observed by the possible participants in the run $\AAA$, before $x \boxplus {\token}$ is known. Besides $s$, known by Victor and Peggy, and $a$, $b$ and $x$, announced publicly but never reused, the context $\crbefo{\AAA}{x \boxplus {\token}}$ thus also contains a single additional challenge $z$, issued by the $\AAA$ttacker, and the corresponding response $z\boxplus {\token}$ (provided by Peggy before she receives Victor's challenge $x$). 

Moreover, the $\AAA$ttacker may issue a 
family $Y\subseteq \ZZz_2^\ell$ of additional challenges to Peggy, and construct a list $\{b_y\}_{y\in Y}$ of the future values of Victor's counter. To each new challenge, Peggy will respond with $y\boxplus \token_y$, where each response token $\token_y= H(s::a_y::b_y)$ is derived using a new value of the counter $a_y$. By assumption, $\indep{\token_y}{\token}$ holds for all $y$. Independently of the distance of $Y$ and the challenge $x$, the responses $y\boxplus \token_y$ will provide no information about $x\boxplus \token$. In summary, the term context $\crbefo{\AAA}{x \boxplus {\token}}$ is thus
\[
 \left\{s,a,b,x,z, z\boxplus {\token}\right\}   \cup 
\{y, a_y, b_y, y\boxplus h_y \ |\ h_y = H(s.a_y.b_y)\wedge y\in Y\} 
\]

\noindent for some $Y\subseteq \ZZz_2^\ell$, where $a:Y\rightarrow \ZZz_2^\ell$ is injective, and $b:Y\rightarrow \ZZz_2^n$ arbitrary. The assumption about $H$ implies $\indep {y,a_y, b_y, y\boxplus h_y\ }{\ x\boxplus {\token}}$, which further tells that for any $\Xi \subseteq \AAA\left(x\boxplus {\token}\right)$ 
\bear
\{s, a, b, z,z\boxplus {\token}\}\cap \Xi =   \emptyset   
& \Longrightarrow & \indep {\Xi\ } {\ x\boxplus {\token}}
\eear
and we are done.

The proof of \ref{sets}{\em (b)} is analogous, but slightly simpler, elaborating the fact that obtaining one challenge tells nothing about another one. 
\epr

\bprf { of Lemma \ref{grindout}} Since ${\token}$ is indistinguishable from random, the bits of any $h_\ell$ are indistinguishable from
independent. The probability of guessing any chosen substring of length $\ell$ in $\token$ is indistinguishable from $2^{-\ell}$. In particular, the probability of guessing $x_\ell \boxplus {\token}_\ell$ for a chosen $x_\ell$ is indistinguishable from $2^{-\ell}$. Knowing which substring is being guessed presents no advantage, and thus $\pder{x_\ell}{x_\ell\boxplus {\token}_\ell} = 2^{-\ell}$. 

Equations \eqref{bbb} and \eqref{ccc} follow from Prop.~\ref{kernel}.
\epr

\bprf{ of Prop.~\ref{guardprop}}
The claim follows from the fact that each set $\Xi\subseteq\crbefo{\AAA}{x \boxplus {\token} }$ such that
$\adv{\Xi}{x \boxplus {\token} } \gt 0$ satisfies at least one of the following inequalities:
\begin{gather}
 \pder{\Xi}{x  \boxplus  {\token}}   \le 
 \pder{\Xi}{s}\cdot \pder{\Xi, s}{x \boxplus {\token}} \hspace{.8em}  \label{auth1} \\
 \pder{\Xi}{x   \boxplus  {\token}}   \le 
 \pder {\Xi}{   z   \boxplus  {\token} }\cdot   \pder{\Xi , z   \boxplus  {\token}  }{ x \boxplus {\token}} \label{auth2} 
\end{gather}
According to Lemma~\ref{sets}(a) for each subset $\Xi$ of $\AAA( x \boxplus {\token})$  such that $a \in \Xi$, it suffices to consider the set $\Xi \cap \left\{s,a,b,x,z,z\boxplus {\token}\right\}$.  
Once the problem is reduced this far, the rest follows by case analysis,  using Lemma~\ref{grindout}.
\epr

\bprf{ of Prop.~\ref{guardprop2}} 
The claim is that each $\Xi\subseteq \crbefo{\EEE}{x}$ such that $\adv{\Xi}{x \boxplus {\token} } \gt 0$ satisfies
\beq\label{fresh}
 \pder{\Xi}{x  \boxplus  {\token}}   \le 
  \pder{\Xi}{\subs{x}{\kernel {\token}}} \cdot 
 \pder%
 {\Xi, \subs{x}{\kernel {\token}}}
 {x \boxplus {\token}}
 \eeq
 Lemma~\ref{sets}(b)  says that it suffices to consider $\Xi \cap \left\{s,a,b\right\}$ if $a \in \Xi$.   Thus, we only need to consider the subsets of  $\left\{s,a,b\right\}$, and since $b$ is deterministic, this reduces to the subsets of
 $\{s,a\}$. The assumption that the stream $h$ is indistinguishable from random implies  $\pder{\Xi}{x_\boxplus {\token}}_\ell = 2^{-\ell}$ whenever $\Xi$ is a proper subset of $\{s ,a\}$. So (\ref{fresh}) holds trivially in that case. For $\Xi = \{s ,  a \}$, using Prop.~\ref{kernel} and Lemma \ref{grindout}, we have $\pder{\Xi}{x \boxplus {\token}}_\ell = \pder{\Xi}{\subs x {\kappa {\token}}}_\ell = 2^{| {\kappa {\token}}|- \ell}$ and on the other hand $\pder{\Xi,\subs x {\kappa {\token}}}{x \boxplus {\token}}_\ell =1$. Hence (\ref{fresh}).
\epr

\bprf{ of Prop.~\ref{34}} Since $\Prob(x\in \ZZz_2^\ell) = 2^{-\ell}$ by assumption, and $\pder{x,z,z\boxplus {\token}}{x\boxplus {\token}} = 2^{-\Delta(z,x)}$ by (\ref{crude}), it follows that
\begin{multline*} 
\sum_{x\in \ZZz_2^\ell} 2^{-\ell}\pder{x,z,z\boxplus {\token}}{x\boxplus {\token}}_\ell = 
 2^{-\ell}\cdot \sum_{i=0}^{\ell}\binom{\ell}{i} 2^{-i}
  =    2^{-\ell}\cdot \frac{3^\ell}{2^\ell}\ = \ \left(\frac{3}{4}\right)^\ell
\end{multline*}
\epr

\bprf{ of Prop.~\ref{early}} 
By hypothesis the token ${\token} = H(s::a::b)$ is indistinguishable from a random value. Since $\indep{s,a,b}{x}$ also holds by assumption, $\pder{s,a,b}{x\boxplus {\token}} = \pder{{\token}}{x\boxplus {\token}}$ follows, because $s,a,b$ can only be useful to derive ${\token} = H(s::a::b)$. But Prop.~\ref{kernel}(a) then implies that $\pder{s, a, b}{x\boxplus \token}_\ell =  2^{i-\ell}$, where $i = \lvert \kappa {\token}\rvert$. 
The expected value that Peggy will guess $x \boxplus {\token}$ are averaged over the possible values of $\token$, and hence
\[\begin{split}
\sum_{\token \in \ZZz_2^{\ell}} \sum_{x \in \ZZz_2^{\ell}}2^{-\ell} \pder{\token}{x \boxplus \token}_\ell & =\ \  
  2^{-\ell} \cdot \sum_{i}^{\ell} \binom{\ell}{i} 2^{i-\ell}
=  2^{-2\ell} \cdot {3^\ell}\ = \ \left( \frac{3}{4} \right)^{\ell}
\end{split}\]
\epr

\bprf{ of Thm.~\ref{mainthm}} By \eqref{un}, to prove the Theorem, it suffices to show that both $\Prob(\AAA|\VVV)$ and $\Prob(\EEE|\VVV)$ are negligible.
The Bayes' Theorem and the hypotheses imply
\bear
\Prob(\AAA |\VVV) & = &  \frac{\Prob(\VVV\ |\ \AAA)\
\cdot\ \Prob(\AAA)}{\Prob(\VVV)}  \le \frac{\Prob(\VVV\ |\ \AAA)\
\cdot\ C}{D}
\eear
\noindent Since $\Prob(\VVV|{\AAA})$ is negligible by Prop.~\ref{34}, $\Prob(\AAA|\VVV)$ is negligible too.
The fact that $\Prob(\EEE|\VVV)$ is negligible follows in the same way from Prop.~\ref{early}.
\epr

\end{document}